# Using Unsupervised Domain Adaptation Semantic Segmentation for Pulmonary Embolism Detection in Computed Tomography Pulmonary Angiogram (CTPA) Images

Wen-Liang Lin [1]*, Yun-Chien Cheng [1]*,

[1] Department of Mechanical Engineering, College of Engineering, National Yang Ming Chiao Tung University, Hsin-Chu, Taiwan

*Corresponding author: lightneil9.en10@nycu.edu.tw, yccheng@nycu.edu.tw

## Abstract

While deep learning has demonstrated considerable promise in computer-aided diagnosis for pulmonary embolism (PE), practical deployment in Computed Tomography Pulmonary Angiography (CTPA) is often hindered by "domain shift" and the prohibitive cost of expert annotations. To address these challenges, an unsupervised domain adaptation (UDA) framework is proposed, utilizing a Transformer backbone and a Mean-Teacher architecture for cross-center semantic segmentation. The primary focus is placed on enhancing pseudo-label reliability by learning deep structural information within the feature space. Specifically, three modules are integrated and designed for this task: (1) a Prototype Alignment (PA) mechanism to reduce category-level distribution discrepancies; (2) Global and Local Contrastive Learning (GLCL) to capture both pixel-level topological relationships and global semantic representations; and (3) an Attention-based Auxiliary Local Prediction (AALP) module designed to reinforce sensitivity to small PE lesions by automatically extracting high-information slices from Transformer attention maps. Experimental validation conducted on cross-center datasets (FUMPE and CAD-PE) demonstrates significant performance gains. In the $FUMPE \rightarrow CAD\text{-}PE$ task, the IoU increased from 0.1152 to 0.4153, while the $CAD\text{-}PE \rightarrow FUMPE$ task saw an improvement from 0.1705 to 0.4302. Furthermore, the proposed method achieved a 69.9% Dice score in the $CT \rightarrow MRI$ cross-modality task on the MMWHS dataset without utilizing any target-domain labels for model selection, confirming its robustness and generalizability for diverse clinical environments.

Keywords: Deep learning, Pulmonary embolism, Vision Transformer, Computed Tomography Pulmonary Angiography, Unsupervised domain adaptation, Semantic segmentation

## 1. Introduction

Pulmonary embolism (PE) is a life-threatening cardiovascular condition requiring prompt diagnosis and immediate clinical intervention. Computed Tomography Pulmonary Angiography (CTPA) is currently regarded as the clinical gold standard for PE diagnosis [1] due to its superior spatial resolution and diagnostic accuracy. However, manual interpretation of CTPA scans remains a formidable challenge for radiologists; PE lesions, particularly those at the sub-segmental level, are often remarkably small and can be easily misidentified or overlooked amidst complex pulmonary vascular structures. To mitigate the risks of misdiagnosis and reduce the workload of clinicians, deep learning-based Computer-Aided Diagnosis (CAD) systems have shown significant progress in automating PE detection. Early studies employing Convolutional Neural Networks (CNNs), such as those by Cano-Espinosa et al. [2] and Long et al. [3], demonstrated that direct segmentation strategies and probability-driven anchor mechanisms could effectively detect sub-segmental emboli within single-center datasets. Similarly, Trongmetheerat et al. [4] integrated attention mechanisms into U-Net architectures to enhance the sensitivity of detecting minute lesions in small vessels.

However, these supervised approaches rely on the Independent and Identically Distributed (IID) assumption. In clinical reality, medical images are subject to significant "domain shift" caused by variations in scanner protocols, contrast timing, and patient demographics [5]. Consequently, models trained on source domains often fail to generalize to unseen target domains (out-of-distribution data). While retraining models on target data is a theoretical solution, the prohibitive cost of pixel-level annotation by radiologists renders fully supervised transfer learning impractical. To address this, Unsupervised Domain Adaptation (UDA) has become a key research direction that seeks to transfer knowledge learned from a labeled source domain to an unlabeled target domain without requiring target-domain annotations [6].

Extensive research has explored various approaches for UDA. These methods can generally be grouped into three categories: image-level alignment, adversarial learning, and self-training. Image-level approaches span from CycleGAN-based translation methods [7-9], which can be computationally prohibitive, to lightweight alternatives such as Fast Fourier Transform (FFT)-based style transfer [10-14]. In parallel, feature- or output-level adversarial alignment [15-17] has been widely explored, yet it is often difficult to train reliably due to the sensitivity of min–max optimization. These limitations have motivated increasing interest in self-training, where Mean-Teacher [18-19] frameworks are frequently adopted for their relative stability. Nevertheless, a central bottleneck remains: pseudo-labels on the target domain are inevitably noisy, leading to error accumulation in highly class-imbalanced scenarios like PE segmentation.

Beyond the choice of adaptation strategy, the backbone architecture also plays a critical role. Many UDA pipelines were originally developed on standardized CNN backbones. However, CNN-based models have increasingly been outperformed by State-of-the-Art (SOTA) Transformer architectures [20-22]. By leveraging self-attention to model long-range semantic dependencies, Transformers can overcome the locality of CNN receptive fields and often exhibit stronger generalization [23-24]. Building on this trend, DAFormer [25] was among the first to integrate a Transformer backbone into a Mean-Teacher UDA framework, achieving SOTA performance on the $GTA \rightarrow Cityscapes$ benchmark. In medical imaging,

MA-UDA [26] further advanced this direction by explicitly aligning representations at both the pixel and attention levels to mitigate cross-modality domain shift. Its key innovation, the Meta Attention mechanism, establishes hierarchical correlations across multi-head attention maps, enabling efficient multi-level alignment without introducing multiple discriminators as required in adversarial approaches.

In the current landscape of medical UDA, MAPSeg [27] represents the peak of non-adversarial SOTA, achieving an impressive 81.32% Dice score on $CT \rightarrow MRI$ tasks. By abandoning unstable adversarial learning for a robust Mean-Teacher mechanism, it successfully integrates anatomical priors via a Global-Local Collaboration (GLC) module. Notably, MAPSeg established a new benchmark by proposing a fair model selection mechanism, addressing the common but clinically impractical practice of relying on target-domain labels for validation. However, despite these breakthroughs, significant barriers remain. First, MAPSeg's reliance on 3D CNNs and 3D VAEs imposes excessive computational overhead and memory demands, raising the threshold for hardware-constrained clinical environments. Second, and most critically for PE detection, the GLC module employs random cropping strategies. Given that PE lesions occupy a minute fraction of the volume, random sampling frequently yields background-only patches, leading to ineffective anatomical alignment and the loss of critical lesion-related context.

To bridge these gaps, this study proposes a Transformer-based Mean-Teacher UDA framework tailored for cross-center PE detection. Unlike traditional CNNs, a Mix Vision Transformer (MiT) backbone [22] is employed to capture long-range semantic dependencies while preserving high-resolution details for small lesions. The primary contribution lies in the integration and design of three feature-space alignment modules to enhance pseudo-label reliability: (1) Prototype Alignment (PA) to minimize category-level distribution shifts; (2) Global and Local Contrastive Learning (GLCL), augmented with a momentum queue, to capture topological relationships and global semantics without large-batch requirements; and (3) an Attention-based Auxiliary Local Prediction (AALP) module. Unlike random cropping, AALP utilizes Transformer attention maps to explicitly extract lesion-rich regions, enforcing anatomical consistency between local and global views.

The main contributions of this study are summarized as follows:

(1) A computationally efficient Transformer-based Mean-Teacher UDA framework is developed for CTPA PE segmentation, balancing performance and resource constraints.
(2) Three feature space alignment modules (PA, GLCL, AALP) are integrated and designed to actively improve pseudo-label quality and address the challenge of detecting small PE lesions.
(3) An Attention-based Auxiliary Local Prediction (AALP) module is proposed to replace random cropping, significantly enhancing sensitivity to tiny objects.
(4) Experiments on cross-center (FUMPE, CAD-PE) and cross-modality (MMWHS) datasets demonstrate that the proposed method effectively mitigates domain shift and achieves robust performance under a strictly unsupervised model selection setting.

## 2. Related Work

### 2.1. Unsupervised Domain Adaptation

UDA aims to minimize the expected risk on the target domain by aligning the source and target distributions, as theoretically bounded by Ben-David et al. [28]. Current strategies fall into three categories:

(1) Image-Level Alignment: These methods reduce visual discrepancies in the input space. While CycleGAN-based approaches [7-9] successfully align textures, they entail high memory usage and training instability due to the requirement of multiple generators and discriminators. Alternatively, Fourier Transform (FFT) based methods [10-14] decompose images into amplitude (style) and phase (structure). By swapping the low-frequency amplitude spectrum of source images with that of target images, style transfer can be achieved efficiently without learnable parameters. Similarly, histogram matching aligns intensity distributions. Given the computational constraints in clinical settings, this study adopts efficient FFT and histogram matching techniques for preliminary input-level alignment.
(2) Feature-Level Adversarial Alignment: Inspired by GANs, methods like AdaptSegNet [15] and CyCADA [16] employ domain discriminators to enforce feature invariance. Although effective, adversarial training is notoriously unstable and sensitive to hyperparameters. Consequently, this study opts for a more stable self-training paradigm.
(3) Self-Training and Mean-Teacher: Self-training generates pseudo-labels for the unlabeled target domain to guide learning [10, 25, 27, 29-37]. The Mean-Teacher framework [18-19], utilizing Temporal Ensembling, has become a standard for stabilizing these pseudo-labels. However, the quality of pseudo-labels remains a bottleneck. While threshold-based filtering and uncertainty estimation [38] are common, they are passive strategies. Recent works have explored meta-learning [39] for noise correction, but computational costs remain high. This study focuses on improving pseudo-labels through active structural learning in the feature space.

**2.2. Structural Learning in Feature Space**

To enhance feature discriminability, Prototype Alignment (PA) and Contrastive Learning have been introduced to UDA. PA aligns class-specific centroids between domains [12], reducing category-level shifts. However, PA alone may fail for class-imbalanced data like PE, where centroids are biased by background noise. Contrastive learning pulls positive pairs closer while pushing negatives apart. Liu et al. [13] proposed a Dual Contrastive framework combining global and local contrastive losses to capture both semantic layout and pixel-level topology. A limitation of such approaches is the dependency on large batch sizes for sufficient negative samples. To mitigate this, this study incorporates a momentum queue (MoCo) [40] into the GLCL module, enabling effective contrastive learning under limited hardware resources.

**2.3. Context-Aware Modeling and Limitations of Random Crop**

Integrating global and local context is vital for ensuring anatomical plausibility. MAPSeg introduced a Global-Local Collaboration (GLC) module that enforces consistency between global features and local patches. While achieving SOTA results, MAPSeg relies on random cropping to select local patches. In PE detection, where lesions are sparse and small, random cropping predominantly selects background regions, rendering the local-global alignment ineffective. To address this critical flaw, this study proposes the Attention-based Auxiliary Local Prediction (AALP) module, which leverages the Transformer's inherent attention maps to intelligently locate and crop lesion-informative regions, ensuring meaningful context alignment.

**3. Material and Methods**

**3.1. Datasets for UDA Training**

To evaluate the generalizability of the proposed UDA framework in clinical scenarios, two types of adaptation tasks were conducted: bidirectional cross-center adaptation using CTPA images and cross-modality adaptation using cardiac scans.

*3.1.1. CTPA Datasets*

For the cross-center adaptation experiments, two publicly available CTPA datasets with expert semantic annotations for PE were utilized: the Ferdowsi University of Mashhad's Pulmonary Embolism (FUMPE) dataset [41] and the Computer Aided Detection for Pulmonary Embolism Challenge (CAD-PE) dataset [42]. Specifically, the experiments in this research were conducted using image sequences from 33 selected patients from the FUMPE dataset and 91 patients from the CAD-PE dataset. A critical aspect of the experimental design is the patient-wise partitioning of data, rather than slice-wise partitioning, to better reflect authentic clinical conditions and prevent data leakage. Specifically, the training and validation sets for FUMPE and CAD-PE were partitioned into 29/4 and 80/11 patients, respectively. In the UDA training phase, the model has access to annotations only from the source domain, while the target domain remains entirely unlabeled. For instance, in the $FUMPE \rightarrow CAD\text{-}PE$ task, the framework learns from the labeled FUMPE training set and performs domain alignment using the unlabeled CAD-PE training set, with performance finally evaluated on the CAD-PE validation set.

*3.1.2. Cardiac CT-MRI Dataset*

To ensure a fair comparison with existing medical UDA literature, we conduct cross-modality adaptation experiments on the Multi-Modality Whole Heart Segmentation (MMWHS) dataset [43]. MMWHS comprises 20 MRI scans and 20 CT scans with annotations for seven cardiac anatomical structures. Following the prevalent standards in UDA research, this study focuses on four key structures: Myocardium of the Left Ventricle (MYO), Left Atrium Blood Cavity (LAC), Left Ventricle Blood Cavity (LVC), and Ascending Aorta (AA). To ensure direct comparability with SOTA methods, we adopt the same training/validation split protocol as specified in [8].

**3.2. Experimental Procedure**

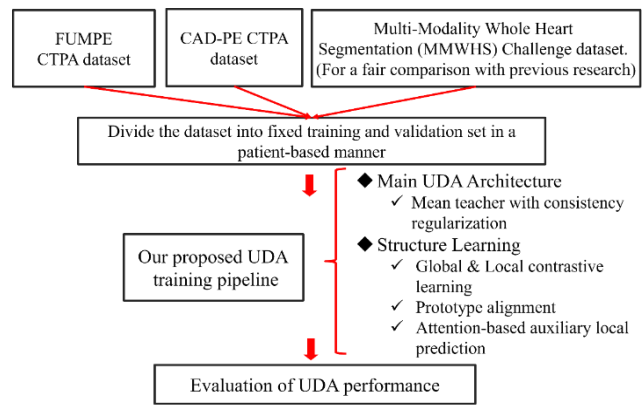

Figure 1. Experiment process

The overall experimental procedure of this study is illustrated in Fig. 1. Following the dataset partitioning strategies detailed in Section 3.1, the proposed UDA training framework is implemented to develop the semantic segmentation model. To comprehensively evaluate the performance and clinical utility of the system, the following tests were conducted:

(1) Validation of the UDA framework's effectiveness on the MMWHS dataset for cross-modality adaptation.
(2) Assessment of the performance and adaptability of the Transformer-based architecture in UDA tasks.
(3) Validation of the UDA framework on CTPA datasets, specifically evaluating the performance gain in detecting minute pulmonary embolism lesions through feature space structural learning.
(4) Ablation studies on each critical UDA component to verify their individual contributions.
(5) A fair comparison between the proposed UDA training framework and existing SOTA methods.

### 3.3. Data Preprocessing

The raw CTPA data is provided in DICOM or Nearly Raw Raster Data (NRRD) format, where intensities are expressed in Hounsfield Units (HU) to reflect tissue radiation absorption. The original HU range, typically spanning from -1000 to 3000, often results in poor contrast between PE lesions and surrounding pulmonary structures. To address this, an empirical HU window is applied to clip the intensity range to [-200, 500], followed by normalization to [0, 1]. Furthermore, as the pulmonary artery is typically situated at the image center, the images are cropped to a resolution of $416 \times 416$ to reduce computational redundancy and overhead. The visual enhancement achieved through this HU windowing is illustrated in Fig. 2. For the MMWHS dataset, the preprocessing protocols strictly adhere to the standards established in [8], including random cropping to a size of $256 \times 256$ to ensure the direct comparability of experimental results.

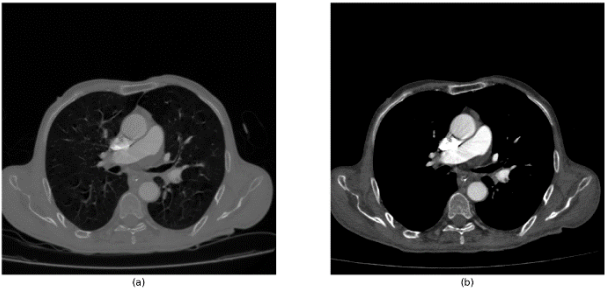

Figure 2. (a) Before applying a HU window (b) After applying a HU window

### 3.4. UDA Training Methods

#### 3.4.1. Problem Definition

Given a source domain image set $X_s = \{x_s^i\}_{i=1}^{N_s}$ with its corresponding pixel-level ground truth labels $Y_s = \{y_s^i\}_{i=1}^{N_s}$, and a related target domain image set $X_t = \{x_t^i\}_{i=1}^{N_t}$, where $N_s$ and $N_t$ represent the total sample counts for each respective domain. Our objective is to construct a robust semantic segmentation model $g_\theta$. Conventional models optimized solely on source-domain data frequently experience a substantial drop in predictive accuracy when applied to the target environment, a consequence of the domain shift effect. UDA aims to mitigate this discrepancy without relying on target-domain annotations. By implementing mechanisms such as image-level translation or feature-space alignment to minimize distributional variances, UDA facilitates the migration of source-learned knowledge to the target domain. This process eventually enhances the model's diagnostic efficacy in specialized clinical target-domain applications.

#### 3.4.2. Overall UDA Training Framework

The conceptual configuration of the developed UDA training framework is presented in Fig. 3, where the green and red lines identify the information flow for the source and target domains, respectively. The overall framework is structured into three essential phases. The process starts with style-transfer data augmentation during the initial processing to mitigate visual discrepancies across the domains. Subsequently, a Mean-Teacher self-training scheme with consistency regularization is employed; the teacher network produces pseudo-labels for unlabeled target images, providing supervisory signals to the student model despite the absence of target-domain annotations. Finally, to further optimize feature representation, three specialized modules for feature space structural learning are integrated into the training process. These modules are designed to mine semantic structures within the latent space, which improves pseudo-label reliability and enhance sensitivity to the fine boundaries of small lesions. Detailed formulations and implementation details are provided in the following subsections.

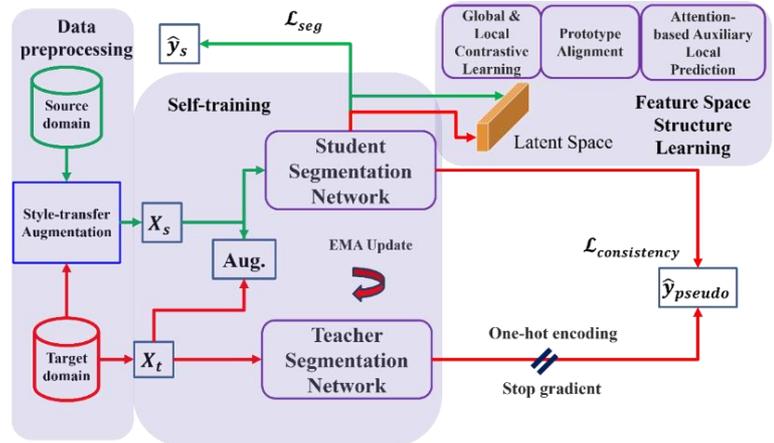

Figure 3. Overall UDA training framework

*3.4.3. Segmentation Network Architecture*

As established in [25], Transformer-based models exhibit superior performance and flexibility over classical CNNs when applied to UDA semantic segmentation. Consequently, the Mix Vision Transformer (MiT-B5), pretrained on ImageNet [44], serves as the encoder in our architecture. The decoder utilizes a Feature Pyramid Network (FPN) following the framework outlined in [45]. The FPN architecture is particularly critical for enhancing the detection of small lesions, as it effectively integrates high-resolution details and rich semantic information from shallow and deep layers, respectively. Finally, these combined feature maps are processed by a segmentation head to produce pixel-level class predictions.

*3.4.4. Style Transfer and Data Augmentation*

In this study, the image sets generated through data augmentation are denoted by $\{X_s, X_{s \to t}, X_{t \to s}, X_t\}$. The sets $\{X_s, X_t\}$ consist of source and target images processed via conventional augmentation strategies, such as Gaussian blurring, rotations, and random pixel dropout. Conversely, $\{X_{s \to t}, X_{t \to s}\}$ denote images processed via the style transfer modules (Fourier-based augmentation or Histogram Matching) directly minimize input-space domain disparities.

**For Fast Fourier Transform (FFT) Data Augmentation:** As established in prior UDA research [10-11], Fourier-based style transfer has demonstrated significant effectiveness in enhancing data diversity and improving performance across various UDA tasks. This technique relies on the decomposition of frequency-domain images into their respective phase and amplitude constituents. Generally, the amplitude component captures stylistic attributes like texture and intensity, whereas the phase component maintains the essential semantic and structural layout of the image. For a source image $X_s \in \mathbb{R}^{H \times W \times C}$ (where $H$, $W$, and $C$ signify height, width, and channel indices), the Fourier transform $\mathcal{F}$ is utilized to obtain the amplitude spectrum $\mathcal{A}_s$ and the phase spectrum $\mathcal{P}_s$ as follows:

$$\mathcal{F}(X_s)(u,v,c) = \sum_{h,w} X_s(h,w,c) e^{-j2\pi\left(\frac{h}{H}u + \frac{w}{W}v\right)} \quad (1)$$

where $(h,w)$ are coordinates in the spatial domain, $(u,v)$ are coordinates in the frequency domain, and $j$ is the imaginary unit ($j^2 = -1$). Similarly, the amplitude $\mathcal{A}_t$ and phase $\mathcal{P}_t$ of a target image $X_t$ are obtained. Subsequently, a binary mask $\mathcal{M}_\beta$ is defined to isolate the central low-frequency components (set to 1) while setting all remaining frequencies to 0, thereby enabling the source image to be rendered in the target style. The area of this region is controlled by the parameter $\beta$. The style transfer operation is formulated as:

$$\mathcal{A}_{s \to t} = \mathcal{A}_s \odot (1 - \mathcal{M}_\beta) + \mathcal{A}_t \odot \mathcal{M}_\beta \quad (2)$$

Here, the symbol $\odot$ signifies the Hadamard (element-wise) product. This strategy ensures the preservation of high-frequency source structure while integrating low-frequency target characteristics, thereby expanding the diversity of the training data. Finally, the hybridized amplitude $\mathcal{A}_{s \to t}$ is combined with the original phase $\mathcal{P}_s$ via the inverse Fourier transform $\mathcal{F}^{-1}$ to reconstruct the stylized image $X_{s \to t}$:

$$X_{s \to t} = \mathcal{F}^{-1}(\mathcal{A}_{s \to t}, \mathcal{P}_s) \quad (3)$$

**For Histogram Matching:** This classical image processing technique maps the intensity distribution of a source image to that of a target image to achieve alignment. For $X_s$ and $X_t$, the pixel intensities are first normalized into $I$ discrete bins within the range $[0, I-1]$. Letting $s_k$ be the intensity value of the $k$-th bin, the probability distribution $p_s(s_k)$ is defined as:

$$p_s(s_k) = \frac{n_k}{HW} \quad (4)$$

where $n_k$ is the number of pixels with intensity $s_k$, and $H \times W$ is the image size. A similar distribution $p_t(t_k)$ is calculated for the target domain. By computing the cumulative distribution functions (CDF) for both domains, a mapping function is established to adjust the intensity distribution of the source image to match the target, resulting in the stylized output $X_{s \to t}$.

*3.4.5. Mean Teacher and Consistency Regularization*

To mitigate the domain shift between the source and target domains, we adopt a self-training strategy based on the Mean-Teacher paradigm [18]. This iterative framework generates online pseudo-labels to supervise learning on unlabeled target images. The framework consists of a teacher model $f(X; \Theta_{teacher})$ and a student model $f(X; \Theta_{student})$, which share identical architectures. Instead of being updated by backpropagation, the teacher's weights are adjusted at every training step $t$ through the Exponential Moving Average (EMA) of the student's weights:

$$\Theta_{teacher}^t = \alpha \Theta_{teacher}^{t-1} + (1 - \alpha) \Theta_{student}^{t-1} \quad (5)$$

where $\alpha$ is a hyperparameter set between 0.99 and 0.999, respectively, according to different training stages. This mechanism is designed to stabilize the optimization process, ensuring the generation of reliable, high-fidelity pseudo-labels as the model converges.

During the pseudo-label generation phase, the teacher model produces a probability map $P_t$ for the target image. While conventional methods often employ a fixed threshold $\tau$ to convert these maps into binary pseudo-labels, the teacher model is prone to noise in the early stages of training. In the context of pulmonary embolism detection, where class imbalance is severe and lesions occupy a minute fraction of the image, a fixed threshold can lead to confirmation bias, as the model tends to classify ambiguous regions as background. To mitigate this, an entropy-based dynamic filtering mechanism is adopted. The entropy for each pixel is calculated as follows:

$$\mathcal{L}_{ent} = - \sum_{c \in Classes} p(c|x_t) \log p(c|x_t) \quad (6)$$

where $p(c|x_t)$ represents the predicted probability of target image $x_t$ belonging to class $c$. By minimizing this entropy, the model is driven to make high-confidence predictions on unlabeled target data. In practice, pixels are ranked by their entropy values, and only the top 80% with the lowest entropy (highest confidence) are retained as valid pseudo-labels for consistency loss calculation. The remaining 20% are masked as unreliable, preventing learning stagnation caused by overly rigid thresholds in early iterations.

Furthermore, a consistency regularization strategy utilizing strong and weak data augmentations is integrated to enhance performance. Specifically, the student and teacher models receive strongly and weakly augmented versions of the input, respectively, and are required to produce consistent predictions. Both the source domain segmentation loss $\mathcal{L}_{seg}$ and the target domain consistency loss $\mathcal{L}_{consistency}$ are calculated using Dice Loss, defined as:

$$\mathcal{L}_{seg} = \mathcal{L}_{consistency} = \mathcal{L}_{dice} = 1 - \frac{2 \sum_{i=1}^{N} p_i y_i}{\sum_{i=1}^{N} p_i + \sum_{i=1}^{N} y_i} \quad (7)$$

where $p_i$ is the predicted probability from the student model and $y_i$ represents the target label. For $\mathcal{L}_{seg}$, $y_i$ is the source-domain ground truth, whereas for $\mathcal{L}_{consistency}$, it denotes the teacher-generated online pseudo-label. Dice Loss is particularly effective for medical image segmentation as it is less sensitive to class imbalance, stabilizing the training process dominated by background pixels.

Finally, while filtering low-confidence pseudo-labels is essential, it may not be sufficient to eliminate all noise. To enable the model to actively learn the structural information of the feature space and mine correlations between domains, this study introduces three specialized feature learning modules: Prototype Alignment (PA), Global and Local Contrastive Learning (GLCL), and Attention-based Auxiliary Local Prediction (AALP). These modules optimize feature representation from different perspectives, thereby improving pseudo-label quality and overall segmentation efficacy.

*3.4.6. Prototype Alignment (PA)*

The Prototype Alignment (PA) mechanism implemented in this study aims to resolve the feature distribution shift commonly encountered in unsupervised domain adaptation. By performing class-level centroid alignment, the framework promotes the clustering of features with identical semantic labels—such as pulmonary vascular structures—within the latent space across both domains.

The procedure starts with the extraction of latent representations $e$ from the penultimate layer of the encoder. For the source and target domains, the model utilizes the source ground truth $y^s$ and the target online pseudo-labels $y^t$ to extract class-specific feature sets $\Phi_c$, respectively. The global category prototype $z_c$ is defined as the centroid of all pixel-wise feature vectors within that category. Notably, as previously discussed, the 20% of target pixels identified as noisy via high entropy values are strictly excluded from the target prototype calculation to ensure representative accuracy:

$$z_c = \frac{1}{|\Phi_c|} \sum_{v \in \Phi_c} e_v \quad (8)$$

where $v$ denotes the spatial index of the flattened feature map, and $e_v$ represents the $D$-dimensional feature vector at that location. Given that pulmonary embolism lesions are extremely sparse, a single training batch may not provide stable category-level information. To mitigate this, a momentum update strategy is employed to maintain the global prototypes:

$$z_c \leftarrow \alpha z_c + (1 - \alpha) z_{c,batch} \quad (9)$$

where $\alpha$ is the momentum coefficient (set to 0.01) and $z_{c,batch}$ is the prototype calculated from the current batch. Finally, a cross-domain global prototype alignment loss is established by reducing the Euclidean distance between the source prototype $z_c^s$ and the target prototype $z_c^t$:

$$\mathcal{L}_{pro} = \sum_{c=1}^{C} ||z_c^s - z_c^t||_2 \quad (10)$$

This optimization compels the model to align the feature centroids of both domains, ensuring robust category-level consistency.

*3.4.7. Global and Local Contrastive Learning (GLCL)*

As illustrated in Fig. 4, this study implements a Global and Local Contrastive Learning (GLCL) module, building upon the Dual Contrastive framework proposed by Liu et al. [13]. The primary objective of GLCL is to force the network to decouple structural semantics from superficial style variations (e.g., CT vs. MRI). Since contrastive learning benefits from abundant negative samples but large batches are memory-intensive, this study incorporates Momentum Contrast (MoCo) [40] in the global branch to increase negative-sample diversity with modest GPU memory cost.

**For Local Contrastive Learning (LCL):** The LCL module focuses on capturing structural information, such as the contours and fine details of lesions. Based on the premise that geometric relationships between neighboring pixels should remain invariant across styles, a Local Projection Head consisting of two convolutional layers is utilized to map latent features into a contrastive space $\mathcal{F} \in R^{S \times S \times K}$, where $S \times S$ and $K$ denote the spatial resolution and projection dimension, respectively. For a feature vector $r^m$ at spatial index $m$ in the source feature map $\mathcal{F}_s$, the positive sample $r_+^m$ is defined as the most similar vector at the corresponding location in the stylized map $\mathcal{F}_{t \to s}$, determined via a cosine similarity matrix. All other spatial locations in $\mathcal{F}_{t \to s}$ are treated as the negative sample set $\{r_-^{mn}\}$. The source and target local contrastive losses, $\mathcal{L}_{\ell c}^s$ and $\mathcal{L}_{\ell c}^t$, are formulated as:

$$\mathcal{L}_{\ell c}^s = \frac{1}{S^2} \sum_m - \log \frac{\exp(r^m \cdot r_+^m / \tau)}{\exp(r^m \cdot r_+^m / \tau) + \sum_n \exp(r^m \cdot r_-^{mn} / \tau)} \quad (11)$$

$$\mathcal{L}_{\ell c}^t = \frac{1}{S^2} \sum_m - \log \frac{\exp(z^m \cdot z_+^m / \tau)}{\exp(z^m \cdot z_+^m / \tau) + \sum_n \exp(z^m \cdot z_-^{mn} / \tau)} \quad (12)$$

where $\tau$ is the temperature parameter. The total local loss $\mathcal{L}_{\ell c}$ is the average of the two: $\mathcal{L}_{\ell c} = \frac{1}{2}(\mathcal{L}_{\ell c}^s + \mathcal{L}_{\ell c}^t)$.

**For Global Contrastive Learning (GCL) with MoCo:** While LCL targets "contours and details," GCL aims to capture the "skeleton and layout" of the image. Features are mapped via a Global Projection Head (Conv + Max Pooling) to a feature vector $f$. To maximize discriminative power without requiring excessive batch sizes, a First-In-First-Out (FIFO) queue is maintained to store historical negative features. To ensure cross-batch consistency, negative samples are generated by the teacher model and stored in the queue. If $N$ denotes the number of source domain images within a single batch, the source global contrastive loss $\mathcal{L}_{gc}^s$ is formulated as:

$$\mathcal{L}_{gc}^s = \frac{1}{N} \sum_{i=1}^N - \log \frac{\exp(f_s^i \cdot f_{s \to t}^i / \tau)}{D_i} \quad (13)$$

$$D_i = \exp(f_s^i \cdot f_{s \to t}^i / \tau) + \sum_{j \neq i} \exp(f_s^i \cdot f_s^j / \tau) + \sum_{k=1}^N \exp(f_s^i \cdot f_{t \to s}^k / \tau) \quad (14)$$

In this formulation, $f_s^i$ serves as the anchor (the feature vector of the $i$-th source image), while $f_{s \to t}^i$ is the corresponding stylized feature, treated as the positive sample. The negative samples consist of $f_s^j (i \neq j)$, representing other source images with different semantics within the same batch, and the set $f_{t \to s}^k$ representing target-to-source stylized features. Similarly, for the target domain, assuming the same number of images $N$, the target global contrastive loss $\mathcal{L}_{gc}^t$ is calculated using the feature sets $\{f_t^i, f_{t \to s}^i, f_t^j, f_{s \to t}^k\}$ as the anchor, positive sample, and negative samples, respectively. The total global loss is the average of the two: $\mathcal{L}_{gc} = \frac{1}{2}(\mathcal{L}_{gc}^s + \mathcal{L}_{gc}^t)$. Finally, the overall contrastive loss is calculated as:

$$\mathcal{L}_{contrast} = \lambda \mathcal{L}_{gc} + (1 - \lambda) \mathcal{L}_{\ell c} \quad (15)$$

where $\lambda$ is set to 0.5. This dual-level approach ensures the model remains robust to global stylistic shifts while maintaining high sensitivity to local pathological structures.

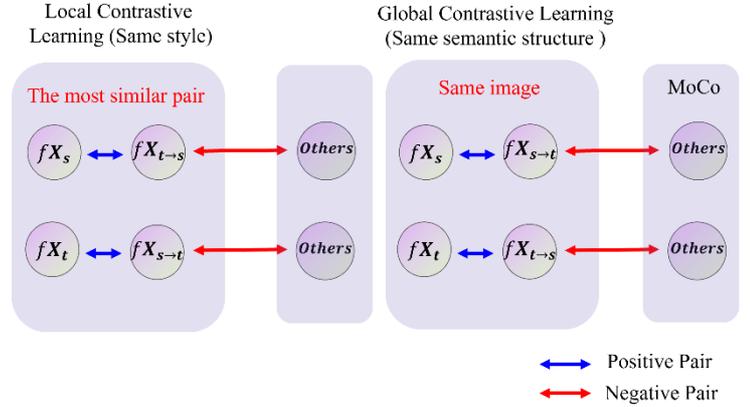

Figure 4. Overview of the GLCL module

*3.4.8. Attention-based Auxiliary Local Prediction (AALP)*

To integrate contextual semantic information, this study adopts the Global-Local Collaboration (GLC) module proposed in [27]. The GLC module facilitates the learning of anatomical priors by aligning global and local predictions, compelling the model to understand long-range spatial relationships rather than over-relying on isolated local features. In clinical scenarios such as PE detection, this constraint is paramount; without global context, a model might erroneously predict lesions or vascular structures in anatomically impossible locations, leading to significant false positives.

The foundational architecture of this integration (previously referred to as GLC) is illustrated in Fig. 5. Let $f(X; \Theta_{student})$ be the student network in the Mean-Teacher framework. The local features $f_{local}$ are derived from a local patch $x$ passing through the student network. For global features, a binary mask $M$ is used to select corresponding spatial regions: $f_{global} = upsample(M \odot g(X))$, where $X$ is the downsampled global image and $g(\cdot)$ represents the feature extraction operation. Subsequently, the local features $f_{local}$ and the masked global features $f_{global}$ are concatenated along the channel dimension. This fused representation is then processed by the decoder $h(\cdot)$ to produce the final segmentation result:

$$D(f) = h(f_{local} \oplus f_{global}) \quad (16)$$

where $\oplus$ denotes the concatenation operation. To prevent local overfitting and ensure anatomical consistency, the distance between $f_{local}$ and $f_{global}$ is minimized using cosine similarity regularization:

$$\mathcal{L}_{cos}(f_{local}, f_{global}) = 1 - \frac{f_{loc} \cdot f_{glo}}{max\left(\|f_{loc}\|_2, \|f_{glo}\|_2\right)} \quad (17)$$

Despite its theoretical advantages, the original GLC module relies on a random cropping strategy to obtain local patches. For CTPA datasets where PE lesions are extremely minute, random cropping frequently yields background-only patches, failing to provide the model with effective lesion-related semantics. To resolve this, this study proposes Attention-based Auxiliary Local Prediction (AALP), which utilizes the Transformer's self-attention mechanism to perform saliency-guided cropping.

The methodology for extracting local patches through the attention mechanism in this study is developed with reference to the approach described in [46]. Specifically, to capture the contribution of various image blocks to the final semantic segmentation, this module leverages the self-attention mechanism of the deep Transformer layers to extract saliency features. We focus on the attention outputs from the final two layers of the model. For a Transformer block with $H$ heads and a $D$-dimensional feature space, the attention matrix $Att_i$ for the $i$-th block is computed via the dot-product of queries ($Q$) and keys ($K$):

$$Att_i = softmax\left(\frac{QK^T}{\sqrt{D/H}}\right) \quad (18)$$

The resulting $(N, N)$ matrix describes the semantic correlation between $N$ image blocks, where the element $Att_{i(j,k)}$ represents the strength of the relationship between the $j$-th and $k$-th blocks. To quantify the specific contribution of each image block, a dimension aggregation operation is performed by fusing features along the second dimension of the matrix to generate a 1D vector, $Attn_{Fuse}$. The final attention saliency map $A$ is then formed by superimposing the $Attn_{Fuse}$ vectors from these final two layers. Blocks with attention values exceeding the mean attention $\bar{a}$ are identified as effective candidate regions. Finally, Connected Component Analysis (CCA) is applied to locate the largest connected area within these regions, which serves as the centroid for cropping the most informative local patch.

The total loss for the AALP module, $\mathcal{L}_{AALP}$, is the sum of the source and target domain losses, $\mathcal{L}_{AALP}^s$ and $\mathcal{L}_{AALP}^t$, respectively:

$$\mathcal{L}_{AALP} = \mathcal{L}_{AALP}^s + \mathcal{L}_{AALP}^t \quad (19)$$

The individual losses for each domain are defined as:

$$\mathcal{L}_{AALP}^S = \gamma \mathcal{L}_{seg}(D(f_s), Y_s) + \delta \mathcal{L}_{cos}(f_{local}^s, f_{global}^s) \quad (20)$$

$$\mathcal{L}_{AALP}^T = 2\gamma \mathcal{L}_{seg}(D(f_t), Y_{pseudo}) + 2\delta \mathcal{L}_{cos}(f_{local}^t, f_{global}^t) \quad (21)$$

where $\mathcal{L}_{seg}$ denotes the Dice Loss, $Y_s$ is the source-domain local patch annotation, and $Y_{pseudo}$ is the target-domain online pseudo-label. Following the established protocols, the hyperparameters are set to $\gamma = 0.05$, $\delta = 0.025$, respectively.

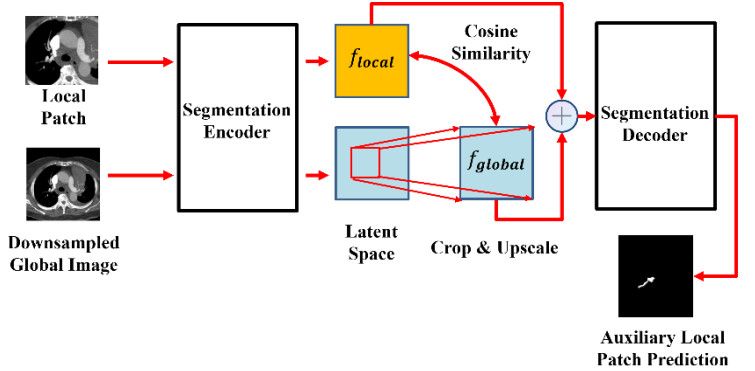

Figure 5. Architecture of the GLC module (redrawn from [27])

### 3.4.9. Overall Objective Function

In summary, the training process of the proposed UDA framework is guided by an overall objective function that integrates the individual contributions of the segmentation, consistency, and structural learning modules. The total loss $\mathcal{L}_{total}$ is formulated as follows:

$$\mathcal{L}_{total} = \lambda_1 \mathcal{L}_{seg} + \lambda_2 \mathcal{L}_{consistency} + \lambda_3 \mathcal{L}_{pro} + \lambda_4 \mathcal{L}_{AALP} + \lambda_5 \mathcal{L}_{contrast} \quad (22)$$

where $\lambda_i$, $i \in \{1,2,\dots,5\}$ denotes the hyperparameters that balance the weight of each loss component during the optimization process.

### 3.5. Implementation Details

The hyperparameters used in this study for UDA

training are $\{\lambda_1, \lambda_2, \lambda_3, \lambda_4, \lambda_5\} = \{1.0, 1.0, 0.1, 1.0, 0.5\}$, and $\beta$ is set to 0.04. The segmentation network was optimized using AdamW, with learning rates of 6e-5 and 6e-4 for the encoder and decoder, respectively, and a weight decay of 5e-4. We adopted cosine annealing for learning-rate scheduling. Models were trained for 120 epochs with a batch size of 8; the first 50 epochs served as a warm-up stage using only source-domain images.

To mitigate the effect of random initialization, each experiment was repeated with five different random seeds, and results are reported as the mean ± standard deviation across runs

### 3.6. Equipment

All experiments were conducted on a workstation featuring an ASUS Z790-A GAMING WIFI 6E motherboard, an Intel Core i9-13900K CPU, and an MSI GeForce RTX 4090 GAMING X TRIO (24 GB) GPU.

## 4. Experimental Results

### 4.1. Evaluation Metrics

To evaluate the segmentation performance of the proposed model on target domain data, Intersection over Union (IoU) and Dice Score are employed as the primary metrics. The calculation formulas are as follows:

$$IoU = \frac{TP}{FP + TP + FN} \quad (23)$$

$$Dice = \frac{2 * TP}{2 * TP + FP + FN} \quad (24)$$

where TP (True Positive) denotes lesion pixels correctly predicted as lesions; FP (False Positive) denotes background/other-class pixels incorrectly predicted as lesions; and FN (False Negative) denotes ground-truth lesion pixels that the model failed to detect.

Regarding the results for the CTPA dataset, IoU is used as the evaluation metric, and only the scores for the pulmonary embolism category are recorded. This is because the pixel distribution between pulmonary embolism lesions and the background in CTPA images exhibits significant class imbalance; including the background class would fail to accurately reflect the model's detection efficacy. Conversely, for the MMWHS cross-modal heart dataset, we follow standard practice and use the Dice score. Dice is computed for each of the selected anatomical structures, and the final performance is reported as the mean Dice across categories.

Finally, for the selection of the optimal model, this study adopts the approach described in [27], which selects the best model without accessing target domain labels. By combining the supervised performance on the source domain with the consistency score between the target domain predictions and pseudo-labels, this strategy simulates the real-world clinical scenario where target domain annotations are unavailable. The selection score is defined as follows:

$$Score = Metric_{source} + 0.5 * Metric_{pseudo} \quad (25)$$

where $Metric_{source}$ represents the performance on the source domain validation set (IoU for the CTPA dataset and Dice Score for the MMWHS dataset, respectively), and $Metric_{pseudo}$ represents the consistency score between the target domain predictions and the generated pseudo-labels (similarly using IoU or Dice Score).

### 4.2. Effectiveness of the Prototype Alignment Module on the MMWHS Dataset (CT → MRI)

This section evaluates the Prototype Alignment (PA) module on the MMWHS dataset (CT → MRI). Experiments utilize an ImageNet-pretrained MiT-B5 encoder, an FPN decoder, and the model selection criteria defined in Section 4.1.

Table 1 shows that the source-only (CT) MiT-B5 model achieves only 19.4% average Dice on the target domain (MRI). Compared to the 84.9% supervised upper bound, this significant gap confirms a severe domain shift. This discrepancy stems from fundamental differences in physical imaging mechanisms, resulting in distinct brightness and texture features that hinder direct model migration.

We define the baseline as a Mean Teacher framework incorporating style transfer (FFT or Histogram Matching) and consistency regularization. Using the FFT-based baseline, performance reaches 62.0%, indicating that FFT effectively aligns input-space styles while Mean Teacher establishes initial decision boundaries via entropy-filtered pseudo-labels.

Integrating the PA module further improves the Dice score to 64.7% (a 2.7% gain), supporting our hypothesis that prototype alignment functions in the deep feature space. By minimizing inter-domain prototype distances, features of the same category cluster more closely. This process effectively corrects noisy pseudo-labels from the Mean Teacher, enhancing cross-modality segmentation precision.

Table 1. Effectiveness of the PA module on the MMWHS dataset (CT → MRI)

\* pretrained on ImageNet　　※ indicates the target labels are not available for validation

| | | Cardiac CT→ Cardiac MRI | |
|---|---|---|---|
| Encoder | Decoder | Method | Average Dice (%) ↑ |
| MiT-B5 * | FPN | Supervised | 84.9 ± 1.0 |
| MiT-B5 * | FPN | W/o adaptation | 19.4 ± 2.4 |
| MiT-B5 * | FPN | Mean teacher with consistency regularization (Baseline) (FFT) | 62.0 ± 1.1 ※ |
| MiT-B5 * | FPN | Baseline (FFT) + Prototype alignment | 64.7 ± 0.7 ※ |

## 4.3. Effectiveness of the Attention-based Auxiliary Local Prediction Module on the MMWHS Dataset (CT → MRI)

This section investigates the effectiveness of the Attention-based Auxiliary Local Prediction (AALP) module on the MMWHS dataset (CT → MRI). To enhance the model's ability to capture anatomical structures, we introduced the local auxiliary prediction module. Table 2 illustrates the impact of different cropping strategies on model performance. Observations show that the local auxiliary prediction module using a random cropping strategy increased the average Dice score from 62.0% to 67.2%, yielding a 5.2% gain. This indicates that learning global and local contextual correlations between features allows the model to effectively capture vital anatomical structural information in medical images. Furthermore, applying cosine constraints to global and local features compels the model to distinguish relationships between neighboring pixels, thereby enhancing the robustness of pseudo-labels against interference.

The proposed Attention-based Auxiliary Local Prediction (AALP) further pushes performance to 68.1%. The key to this additional 0.9% gain lies in "precise positioning." As described in the Methodology chapter, we utilize the Transformer's attention matrix to calculate the global mean attention $\bar{a}$, automatically locating high-attention regions (such as the main heart structure). This ensures that every local patch fed into the auxiliary network contains rich semantic information, avoiding interference from background noise and allowing the model to focus on optimizing the edge details of anatomical structures.

Table 2. Effectiveness of the AALP module on the MMWHS dataset (CT → MRI)

\* pretrained on ImageNet    ※ indicates the target labels are not available for validation

| Cardiac CT→ Cardiac MRI | | | |
|---|---|---|---|
| Encoder | Decoder | Method | Average Dice (%) ↑ |
| MiT-B5 * | FPN | Supervised | 84.9 ± 1.0 |
| MiT-B5 * | FPN | W/o adaptation | 19.4 ± 2.4 |
| MiT-B5 * | FPN | Baseline (FFT) + Auxiliary local prediction (Random Crop) | 67.2 ± 0.3 ※ |
| MiT-B5 * | FPN | Baseline (FFT) + Attention-based auxiliary local prediction | 68.1 ± 0.4 ※ |

## 4.4. Effectiveness of the Global and Local Contrastive Learning Module on the MMWHS Dataset (CT → MRI)

This section investigates the effectiveness of the Global and Local Contrastive Learning (GLCL) module on the MMWHS dataset (CT → MRI). The results are summarized in Table 3.

First, this study explores the impact of different style transfer data augmentations on contrastive learning. Experimental results show that the Dice score using Histogram Matching is 0.6% lower than that using FFT. This discrepancy is likely because Histogram Matching forcibly performs grayscale mapping through a cumulative distribution function (CDF) to align with the target domain. This process often induces a "Staircase Effect" at points of pixel intensity discontinuity. In medical imaging, such digitization distortions can form artifacts similar to actual lesions, negatively affecting the discriminative power of features in contrastive learning.

In terms of contrastive strategies, introducing Global Contrastive Learning (GCL) alone improves the Dice score to 69.2%. GCL forces the network to ignore surface-level style deviations and focus on capturing the global structure and semantic information by pulling together images with "the same semantics but different styles." While this allows the network to understand the spatial layout of anatomical structures from a global perspective and more accurately locate the overall extent of organs, GCL often projects the entire image into a single global feature, which tends to lose pixel-level information. In contrast, Local Contrastive Learning (LCL) aligns pixels of the same style, enabling the model to understand the relationship between each pixel and its neighbors. This leads to more accurate organ boundary delineation and improves the contour clarity of small objects. The results confirm that combining GCL and LCL yields a complementary effect, further increasing the Dice score from 69.2% to 69.5%.

Finally, the GCL in this experiment utilizes MoCo technology, which uses a queue to store historical negative samples. This approach maintains stable feature learning quality and model performance while significantly reducing the requirements for hardware computational resources and GPU memory.

Table 3. Effectiveness of the GLCL module on the MMWHS dataset (CT → MRI)

\* pretrained on ImageNet    ※ indicates the target labels are not available for validation

| Cardiac CT→ Cardiac MRI | | | |
|---|---|---|---|
| Encoder | Decoder | Method | Average Dice (%) ↑ |
| MiT-B5 * | FPN | Supervised | 84.9 ± 1.0 |
| MiT-B5 * | FPN | W/o adaptation | 19.4 ± 2.4 |
| MiT-B5 * | FPN | Baseline (FFT) + global contrastive learning + MoCo | 69.2 ± 0.1 ※ |
| MiT-B5 * | FPN | Baseline (Histogram matching) + global & local contrastive learning + MoCo | 68.9 ± 0.2 ※ |
| MiT-B5 * | FPN | Baseline (FFT) + global & local contrastive learning + MoCo | 69.5 ± 0.2 ※ |

## 4.5. Ablation Study of Key Modules on the MMWHS Dataset

To quantify the contribution of each proposed module to cross-modality adaptation and to verify their complementarity, we conducted a step-by-step ablation experiment on the MMWHS dataset. Table 4. presents the average Dice score performance under different component configurations.

First, using Mean Teacher as the baseline architecture, the Dice scores are 62.0% for the CT → MRI task and 70.3% for the MRI → CT task. After introducing the Prototype Alignment (PA) module, performance improved to 64.7% (+2.7%) and 72.1% (+1.8%) respectively. This result confirms that performing initial category centroid alignment in the feature space effectively corrects the confirmation bias generated by the Mean Teacher.

Significant improvements were observed upon further integrating the Attention-based Auxiliary Local Prediction (AALP). In the CT → MRI task, the Dice score jumped from 64.7% to 69.3%. In the MRI → CT task, the increase was even more substantial, soaring from 72.1% to 79.5%. This indicates that global feature alignment (PA) alone is insufficient to capture fine anatomical boundaries. The AALP module utilizes the attention mechanism for precise localization of the main heart structure, forcing the model to learn high-resolution semantic features within local patches, thereby solving the detail loss common in cross-modality transitions.

Finally, the addition of Global and Local Contrastive Learning (GLCL) served as the ultimate optimization. The model achieved optimal performance of 69.9% and 80.2% in the two tasks, respectively. By pulling similar features closer and pushing dissimilar features apart, GLCL further eliminates boundary ambiguity, resulting in segmentation results with more complete geometric structures.

Table 4. Ablation study of key modules on the MMWHS dataset

※ indicates the target labels are not available for validation

| MT | PA | AALP | GLCL | Average Dice (%) ↑ |
|---|---|---|---|---|
| Cardiac CT → Cardiac MRI | | | | |
| Supervised | | | | 84.9 ± 1.0 |
| W/o adaptation | | | | 19.4 ± 2.4 |
| ✓ | | | | 62.0 ± 1.1 ※ |
| ✓ | ✓ | | | 64.7 ± 0.7 ※ |
| ✓ | ✓ | ✓ | | 69.3 ± 0.1 ※ |
| ✓ | ✓ | ✓ | ✓ | **69.9 ± 0.1** ※ |
| Cardiac MRI → Cardiac CT | | | | |
| Supervised | | | | 88.6 ± 0.7 |
| W/o adaptation | | | | 24.0 ± 1.0 |
| ✓ | | | | 70.3 ± 0.4 ※ |
| ✓ | ✓ | | | 72.1 ± 0.2 ※ |
| ✓ | ✓ | ✓ | | 79.5 ± 0.4 ※ |
| ✓ | ✓ | ✓ | ✓ | **80.2 ± 0.3** ※ |

**4.6. Effectiveness of the Proposed UDA Method on the PE Dataset**

To validate the performance of the proposed UDA framework in real-world clinical scenarios, this study conducted bidirectional cross-center adaptation experiments between two pulmonary embolism datasets from different source centers, FUMPE and CAD-PE. Table 5 presents the performance of the model in terms of the IoU metric.

First, observing the baseline performance without any adaptation (W/o adaptation), when the model is trained on FUMPE and directly inferred on CAD-PE, the IoU is only 0.1152. Conversely, when trained on CAD-PE and inferred on FUMPE, the IoU is only 0.1705. Compared to the theoretical upper bounds of supervised learning (FUMPE: 0.5208 / CAD-PE: 0.4895), this significant performance degradation confirms that even within the same modality (CTPA), significant Domain Shift still occurs due to differences in contrast injection timing, scanning parameters, and patient populations between different hospitals, preventing the model from direct migration.

Upon introducing the UDA method proposed in this study, which includes PA, AALP, and GLCL, the model demonstrated remarkable adaptation capabilities. For FUMPE → CAD-PE, the IoU significantly increased from 0.1152 to 0.4153, bringing a gain of 0.3001. For CAD-PE → FUMPE, the IoU increased from 0.1705 to 0.4302, an increase of 0.2597. These results indicate that our method successfully reduces the distribution gap between cross-center images, proving that the model has successfully established correct lesion feature representations on the unlabeled target domain.

Table 5. Effectiveness of the proposed UDA method on the PE dataset

\* pretrained on ImageNet　※ indicates the target labels are not available for validation

| Encoder | Decoder | Dataset | IoU score on FUMPE validation set | IoU score on CAD-PE validation set | Our proposed UDA Method |
|---|---|---|---|---|---|
| MiT-B5 * | FPN | Trained on FUMPE dataset | 0.5208 ± 0.0086 (Upper bound of below) | **0.1152 ± 0.0117** | **0.4153 ± 0.0072** ※ (+0.3001) |
| MiT-B5 * | FPN | Trained on CAD-PE dataset | **0.1705 ± 0.0127** | 0.4895 ± 0.0121 (Upper bound of above) | **0.4302 ± 0.0094** ※ (+0.2597) |

**4.7. Visualization of Attention-based Local Patch Extraction**

To verify whether the Attention-based Auxiliary Local Prediction (AALP) module proposed in this study can effectively replace random cropping, Fig. 6 showcases examples of local patches localized and cropped by the AALP module based on the attention matrix. The green areas in the figure indicate the ground truth locations of the pulmonary embolism lesions.

In traditional random cropping strategies, because pulmonary embolism lesions account for a very low proportion of the entire CTPA image, randomly selected patches have a high probability of containing only non-informative background. This makes it difficult for the auxiliary network to learn positive sample features.

In contrast to these traditional methods, the results in Figure 6 demonstrate that almost every local patch extracted under the guidance of the attention mechanism contains pulmonary embolism lesions. This confirms that our strategy of utilizing attention weights for localization can effectively filter out low-information background regions, ensuring that every piece of data fed into the auxiliary network possesses learnable information.

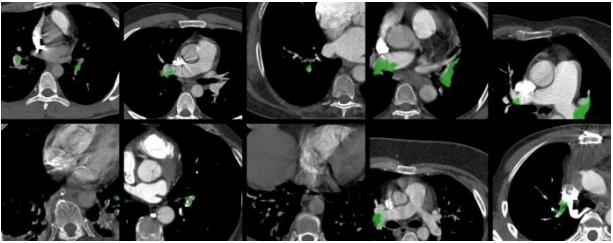

Figure 6. Attention-based local patch samples

### 4.8. Comparison with State-of-the-Art Methods

Table 6 compares the proposed method with current mainstream SOTA methods on the MMWHS dataset (CT → MRI). It is noteworthy that the experimental setup in this study adopts more stringent validation criteria (indicated by ※ in the table), where no target domain labels are used during the model selection and validation phases. Even without the assistance of labels to select the optimal weights, our method achieves a SOTA level of 69.9%, demonstrating high reliability in real-world clinical scenarios with unlabeled data.

Although Table 6 shows that MAPSeg (80.3%) and FSUDA-V2 (75.5%) achieve higher absolute scores than this study, the computational complexity and hardware resource thresholds behind these models must be considered. FSUDA-V2 utilizes a complex ensemble strategy that requires pre-training multiple teacher models and combining contrastive learning to distill knowledge to a student model, resulting in a cumbersome training process and exponentially increased parameter counts and memory requirements. MAPSeg is based on a 3D CNN architecture, which inherently possesses much higher computational demands than the 2D architecture used in this study. Furthermore, it requires a Variational Auto-Encoder (VAE) pre-trained on large-scale datasets to extract prior features, which not only significantly increases training time but also imposes extremely strict requirements on hardware computing power.

Regarding MA-UDA (68.7%), as a representative of adversarial learning, its training process is highly unstable, and literature indicates that it requires high-end NVIDIA V100-level GPUs for training, which presents a significant hardware burden for general clinical laboratories. In contrast, the method proposed in this study requires only a single encoder-decoder architecture paired with efficient feature-space structural learning. Under the hardware condition of a single NVIDIA GeForce RTX 4090 (24 GB VRAM), we achieved high-precision segmentation near 70% based on a 2D network. This indicates that this study successfully achieved a balance between UDA segmentation performance and computational cost, providing a solution better suited for deployment in resource-constrained medical scenarios.

Table 6. Comparison with SOTA methods

※ indicates the target labels are not available for validation

| | Cardiac CT→ Cardiac MRI | |
|---|---|---|
| | Method | Average Dice (%) ↑ |
| Adversarial training | CyCADA [16] (2D – CNN Network) | 57.5 |
| | SIFA-V1 [7] (2D – CNN Network) | 62.1 |
| | SIFA-V2 [8] (2D – CNN Network) | 63.4 |
| | SE-ASA [12] (2D – CNN Network) | 69.9 |
| | MA-UDA [26] (Transformer – based Network) | 68.7 |
| Self-training | DAFormer [25](Transformer – based Network) | 65.9 |
| | FSUDA-V1 [14] (Transformer – based Network) | 70.6 |
| | FSUDA-V2 [13] (Transformer – based Network) | 75.5 |
| | MAPSeg [27](3D – CNN Network) | 80.3※ |
| | Ours | 69.9※ |

## 5. Conclusions

This study presents a Transformer-based UDA framework integrating Prototype Alignment (PA), Attention-based Auxiliary Local Prediction (AALP), and Global/Local Contrastive Learning (GLCL) to mitigate domain shift in medical imaging. Experimental results show that the proposed multi-level alignment significantly improves performance: the MMWHS (CT → MRI) Dice score increased from 19.4% to 69.9%, while the PE cross-center IoU demonstrated substantial bidirectional gains, rising from 0.1152 to 0.4153 (+0.3001) for FUMPE → CAD-PE and from 0.1705 to 0.4302 (+0.2597) for CAD-PE → FUMPE. Notably, the AALP module effectively localizes minute lesions via self-attention, surpassing traditional random cropping. Compared to SOTA methods, our 2D architecture achieves competitive accuracy using only a single RTX 4090, offering a cost-effective and robust solution for real-world clinical deployment

## References

[1] M. T. Lu et al., "Axial and reformatted four-chamber right ventricle–to–left ventricle diameter ratios on pulmonary ct angiography as predictors of death after acute pulmonary embolism," American Journal of Roentgenology, vol. 198, no. 6, pp. 1353-1360, 2012.
[2] C. Cano-Espinosa, M. Cazorla, and G. González, "Computer aided detection of pulmonary embolism using multi-slice multi-axial segmentation," Applied Sciences, vol. 10, no. 8, p. 2945, 2020.
[3] K. Long et al., "Probability-based Mask R-CNN for pulmonary embolism detection," Neurocomputing, vol. 422, pp. 345-353, 2021.
[4] T. Trongmetheerat, K. Sukprasert, K. Netiwongsanon, T. Leeboonngam, and K. Sumetpipat,

"Segment-based and Patient-based Segmentation of CTPA Image in Pulmonary Embolism using CBAM ResU-Net," in Proceedings of the 13th International Conference on Advances in Information Technology, 2023, pp. 1-7.

[5] H. Guan and M. Liu, "Domain adaptation for medical image analysis: a survey," IEEE Transactions on Biomedical Engineering, vol. 69, no. 3, pp. 1173-1185, 2021.

[6] X. Liu et al., "Deep unsupervised domain adaptation: A review of recent advances and perspectives," APSIPA Transactions on Signal and Information Processing, vol. 11, no. 1, 2022.

[7] C. Chen, Q. Dou, H. Chen, J. Qin, and P.-A. Heng, "Synergistic image and feature adaptation: Towards cross-modality domain adaptation for medical image segmentation," in Proceedings of the AAAI conference on artificial intelligence, 2019, vol. 33, no. 01, pp. 865-872.

[8] C. Chen, Q. Dou, H. Chen, J. Qin, and P. A. Heng, "Unsupervised bidirectional cross-modality adaptation via deeply synergistic image and feature alignment for medical image segmentation," IEEE transactions on medical imaging, vol. 39, no. 7, pp. 2494-2505, 2020.

[9] W. Ji and A. C. Chung, "Unsupervised domain adaptation for medical image segmentation using transformer with meta attention," IEEE Transactions on Medical Imaging, vol. 43, no. 2, pp. 820-831, 2023.

[10] Y. Yang and S. Soatto, "Fda: Fourier domain adaptation for semantic segmentation," in Proceedings of the IEEE/CVF conference on computer vision and pattern recognition, 2020, pp. 4085-4095.

[11] Z. Zhou, L. Qi, and Y. Shi, "Generalizable medical image segmentation via random amplitude mixup and domain-specific image restoration," in European Conference on Computer Vision, 2022: Springer, pp. 420-436.

[12] W. Feng, L. Ju, L. Wang, K. Song, X. Zhao, and Z. Ge, "Unsupervised domain adaptation for medical image segmentation by selective entropy constraints and adaptive semantic alignment," in Proceedings of the AAAI Conference on Artificial Intelligence, 2023, vol. 37, no. 1, pp. 623-631.

[13] S. Liu, S. Yin, L. Qu, M. Wang, and Z. Song, "A structure-aware framework of unsupervised cross-modality domain adaptation via frequency and spatial knowledge distillation," IEEE Transactions on Medical Imaging, vol. 42, no. 12, pp. 3919-3931, 2023.

[14] S. Liu, S. Yin, L. Qu, and M. Wang, "Reducing domain gap in frequency and spatial domain for cross-modality domain adaptation on medical image segmentation," in Proceedings of the AAAI conference on artificial intelligence, 2023, vol. 37, no. 2, pp. 1719-1727.

[15] Y.-H. Tsai, W.-C. Hung, S. Schulter, K. Sohn, M.-H. Yang, and M. Chandraker, "Learning to adapt structured output space for semantic segmentation," in Proceedings of the IEEE conference on computer vision and pattern recognition, 2018, pp. 7472-7481.

[16] J. Hoffman et al., "Cycada: Cycle-consistent adversarial domain adaptation," in International conference on machine learning, 2018: Pmlr, pp. 1989-1998.

[17] X. Lai et al., "DecoupleNet: Decoupled network for domain adaptive semantic segmentation," in European Conference on Computer Vision, 2022: Springer, pp. 369-387.

[18] A. Tarvainen and H. Valpola, "Mean teachers are better role models: Weight-averaged consistency targets improve semi-supervised deep learning results," Advances in neural information processing systems, vol. 30, 2017.

[19] K. Sohn et al., "Fixmatch: Simplifying semi-supervised learning with consistency and confidence," Advances in neural information processing systems, vol. 33, pp. 596-608, 2020.

[20] A. Dosovitskiy et al., "An image is worth 16x16 words: Transformers for image recognition at scale," arXiv preprint arXiv:2010.11929, 2020.

[21] Z. Liu et al., "Swin transformer: Hierarchical vision transformer using shifted windows," in Proceedings of the IEEE/CVF international conference on computer vision, 2021, pp. 10012-10022.

[22] E. Xie, W. Wang, Z. Yu, A. Anandkumar, J. M. Alvarez, and P. Luo, "SegFormer: Simple and efficient design for semantic segmentation with transformers," Advances in Neural Information Processing Systems, vol. 34, pp. 12077-12090, 2021.

[23] M. M. Naseer, K. Ranasinghe, S. H. Khan, M. Hayat, F. Shahbaz Khan, and M.-H. Yang, "Intriguing properties of vision transformers," Advances in Neural Information Processing Systems, vol. 34, pp. 23296-23308, 2021.

[24] S. Paul and P.-Y. Chen, "Vision transformers are robust learners," in Proceedings of the AAAI conference on Artificial Intelligence, 2022, vol. 36, no. 2, pp. 2071-2081.

[25] L. Hoyer, D. Dai, and L. Van Gool, "Daformer: Improving network architectures and training strategies for domain-adaptive semantic segmentation," in Proceedings of the IEEE/CVF Conference on Computer Vision and Pattern Recognition, 2022, pp. 9924-9935.

[26] W. Ji and A. C. Chung, "Unsupervised domain adaptation for medical image segmentation using transformer with meta attention," IEEE Transactions on Medical Imaging, vol. 43, no. 2, pp. 820-831, 2023.

[27] X. Zhang et al., "MAPSeg: Unified Unsupervised Domain Adaptation for Heterogeneous Medical Image Segmentation Based on 3D Masked Autoencoding and Pseudo-Labeling," in Proceedings of the IEEE/CVF Conference on Computer Vision and Pattern Recognition, 2024, pp. 5851-5862.

[28] S. Ben-David, J. Blitzer, K. Crammer, A. Kulesza, F. Pereira, and J. W. Vaughan, "A theory of learning from different domains," Machine learning, vol. 79, no.

1, pp. 151-175, 2010.

[29] H. Ma, X. Lin, Z. Wu, and Y. Yu, "Coarse-to-fine domain adaptive semantic segmentation with photometric alignment and category-center regularization," in Proceedings of the IEEE/CVF conference on computer vision and pattern recognition, 2021, pp. 4051-4060.

[30] T. Lei, D. Zhang, X. Du, X. Wang, Y. Wan, and A. K. Nandi, "Semi-supervised medical image segmentation using adversarial consistency learning and dynamic convolution network," IEEE transactions on medical imaging, vol. 42, no. 5, pp. 1265-1277, 2022.

[31] Z. Zhao, F. Zhou, K. Xu, Z. Zeng, C. Guan, and S. K. Zhou, "LE-UDA: Label-efficient unsupervised domain adaptation for medical image segmentation," IEEE Transactions on Medical Imaging, vol. 42, no. 3, pp. 633-646, 2022.

[32] F. Yu, M. Zhang, H. Dong, S. Hu, B. Dong, and L. Zhang, "Dast: Unsupervised domain adaptation in semantic segmentation based on discriminator attention and self-training," in Proceedings of the AAAI Conference on Artificial Intelligence, 2021, vol. 35, no. 12, pp. 10754-10762.

[33] C. S. Perone, P. Ballester, R. C. Barros, and J. Cohen-Adad, "Unsupervised domain adaptation for medical imaging segmentation with self-ensembling," NeuroImage, vol. 194, pp. 1-11, 2019.

[34] L. Hoyer, D. Dai, and L. Van Gool, "HRDA: Context-aware high-resolution domain-adaptive semantic segmentation," in European Conference on Computer Vision, 2022: Springer, pp. 372-391.

[35] H. Wu, Z. Wang, Y. Song, L. Yang, and J. Qin, "Cross-patch dense contrastive learning for semi-supervised segmentation of cellular nuclei in histopathologic images," in Proceedings of the IEEE/CVF conference on computer vision and pattern recognition, 2022, pp. 11666-11675.

[36] H. Yao, X. Hu, and X. Li, "Enhancing pseudo label quality for semi-supervised domain-generalized medical image segmentation," in Proceedings of the AAAI Conference on Artificial Intelligence, 2022, vol. 36, no. 3, pp. 3099-3107.

[37] M. Chen, Z. Zheng, Y. Yang, and T.-S. Chua, "Pipa: Pixel-and patch-wise self-supervised learning for domain adaptative semantic segmentation," in Proceedings of the 31st ACM International Conference on Multimedia, 2023, pp. 1905-1914.

[38] Y. Wang et al., "Semi-supervised semantic segmentation using unreliable pseudo-labels," in Proceedings of the IEEE/CVF conference on computer vision and pattern recognition, 2022, pp. 4248-4257.

[39] X. Guo, C. Yang, B. Li, and Y. Yuan, "Metacorrection: Domain-aware meta loss correction for unsupervised domain adaptation in semantic segmentation," in Proceedings of the IEEE/CVF conference on computer vision and pattern recognition, 2021, pp. 3927-3936.

[40] K. He, H. Fan, Y. Wu, S. Xie, and R. Girshick, "Momentum contrast for unsupervised visual representation learning," in Proceedings of the IEEE/CVF conference on computer vision and pattern recognition, 2020, pp. 9729-9738.

[41] M. Masoudi, H.-R. Pourreza, M. Saadatmand-Tarzjan, N. Eftekhari, F. S. Zargar, and M. P. Rad, "A new dataset of computed-tomography angiography images for computer-aided detection of pulmonary embolism," Scientific data, vol. 5, no. 1, pp. 1-9, 2018.

[42] G. González et al., "Computer aided detection for pulmonary embolism challenge (CAD-PE)," arXiv preprint arXiv:2003.13440, 2020.

[43] X. Zhuang and J. Shen, "Multi-scale patch and multi-modality atlases for whole heart segmentation of MRI," Medical image analysis, vol. 31, pp. 77-87, 2016.

[44] J. Deng, W. Dong, R. Socher, L.-J. Li, K. Li, and L. Fei-Fei, "Imagenet: A large-scale hierarchical image database," in 2009 IEEE conference on computer vision and pattern recognition, 2009: Ieee, pp. 248-255.

[45] A. Kirillov, K. He, R. Girshick, and P. Dollár, "A unified architecture for instance and semantic segmentation," in CVPR, 2017.

[46] S. Zheng, G. Wang, Y. Yuan, and S. Huang, "Fine-grained image classification based on TinyVit object location and graph convolution network," Journal of Visual Communication and Image Representation, vol. 100, p. 104120, 2024.